\documentclass[11pt]{wlscirep}
\usepackage[T1]{fontenc}
\usepackage{float}
\usepackage{bbding}
\usepackage{colortbl}
\usepackage{xcolor}
\usepackage{hyperref}

\newcommand\blfootnote[1]{
    \begingroup
    \renewcommand\thefootnote{}\footnote{#1}
    \addtocounter{footnote}{-1}
    \endgroup
}

\title{Enhanced superconducting properties of Bi$_2$Sr$_2$CaCu$_2$O$_{8+x}$ films with sub-50-nm thickness}

\author[{1,3}]{Bernd Aichner}
\author[{2,3}]{Sandra Keppert}
\author[{2}]{Johannes D. Pedarnig}
\author[{1},\Envelope]{Wolfgang~Lang}
\affil[1]{Faculty of Physics, University of Vienna, Vienna 1090, Austria}
\affil[2]{Institute of Applied Physics, Johannes Kepler University Linz, Linz 4040, Austria}

\affil[3]{these authors contributed equally to this work}

\affil[\Envelope]{email: wolfgang.lang@univie.ac.at}

\begin{abstract}
{\fontfamily{cmss}\fontseries{b}\selectfont Few-unit cell thick Bi$_2$Sr$_2$CaCu$_2$O$_{8+x}$ (Bi-2212) layers have recently attracted much interest due to their extreme anisotropy and two-dimensional superconductivity, although they are typically susceptible to ambient conditions. In this study, we report on thin films approximately 13 unit cells thick that are stable in air, exhibit high anisotropy, and demonstrate extraordinarily high critical currents. By examining the superconducting transition under magnetic fields applied in both out-of-plane and in-plane orientations, we estimate key parameters such as pinning potentials, coherence lengths, London penetration depth, anisotropy factor, and the Ginzburg-Landau parameter. The volume pinning force is better described by a model incorporating an exponential decay term attributed to pronounced thermally-assisted flux flow. The Hall effect in the Bi-2212 films exhibits an extensive anomaly with a double sign change that may challenge existing theoretical explanations for this poorly understood phenomenon in copper-oxide superconductors.}
\end{abstract}

\begin{document}

\flushbottom
\maketitle
\renewcommand{\figurename}{Fig.}

\thispagestyle{empty}

{\flushleft {\fontfamily{cmss}\fontseries{b}\selectfont Keywords} Bi$_2$Sr$_2$CaCu$_2$O$_{8+x}$, High critical currents, Coherence lengths, High anisotropy, London penetration depth, Hall effect anomaly}\blfootnote{\flushleft The version of record of this publication is available free of charge at\\ Scientific Reports {\bf 15}, 11855 (2025); \href{https://doi.org/10.1038/s41598-025-95932-9}{https://doi.org/10.1038/s41598-025-95932-9}}
\vskip 10 pt

\noindent The investigation of high-temperature superconductivity in copper-oxide materials remains one of the most captivating subjects in condensed matter physics. One member of the family of these superconductors, the compound Bi$_2$Sr$_2$CaCu$_2$O$_{8+x}$ (Bi-2212), stands out due to its weakly coupled layered structure that gives rise to the highest anisotropy in this class of materials. The unit cell of Bi-2212 is composed of two diagonally shifted subcells, each incorporating the nominal chemical formula. These distinctive features make it an exemplary system for exploring quasi-two-dimensional superconductivity.

Several studies on Bi-2212 single crystals have exploited their giant anisotropy  \cite{SAMO93,LATY95,HEIN99,WANG01a,BEND07}, revealing novel observations such as the intrinsic Josephson effect along the $c$ direction of the unit cell \cite{KLEI92,OZYU07,KLEI21} and the modified version of the dc flux transformer geometry that confirmed that superconducting vortices are effectively two dimensional, commonly referred to as `pancake vortices' \cite{SAFA92b,BUSC92}.

Experiments with thin Bi-2212 films, epitaxially grown on a suitable substrate, have enabled the realization of well-defined geometries but have been mainly limited to thicknesses of the superconducting layer of about 100\,nm or more \cite{SCHM90,RI94,KONS99,KONS99a,KONS00,ROSS01}. More recently, Bi-2212 whiskers and films with thicknesses of only a few unit cells attracted considerable interest for their enhancement of unconventional Hall effect \cite{ZHAO19}, twisted van-der-Waals \cite{ZHAO23} and surface \cite{WEI23} Josephson junctions, and studies of the Nernst effect \cite{HU24}. Remarkably, the most significant Nernst effect caused by the entropy transport of superconducting vortices was found in a 40-nm-thick Bi-2212 flake, surpassing the respective magnitudes in few-unit cell thin flakes and bulk materials \cite{SHOK24}. While commercially available superconducting-nanowire single-photon detectors (SNSPDs) are based on metallic superconductors, initial success has been reported in implementing them with thin Bi-2212 flakes {\cite{MERI23,CHAR23}}. Extending these concepts to very thin films with larger areas and easier fabrication options may be an interesting future development. 

Few-unit-cell thick Bi-2212 whiskers and monolayers suffer from a major drawback---their instability under ambient conditions \cite{YU19a}. Maintaining their integrity requires sophisticated production processes in an inert gas atmosphere and meticulous handling during transfer to measurement systems \cite{ZHAO23,SHOK24}. In contrast, previously investigated, about 100-nm-thin, Bi-2212 films are stable in dry air and can be handled conveniently. However, they are too thick for applications such as superconducting nanowire single-photon detectors and other experiments where penetration depth is a critical concern. Notably, it has recently been demonstrated that nanometric vortex-pinning landscapes \cite{LANG09,PEDA10,HAAG14} can be patterned with focused He-ion beams into thin YBa$_{2}$Cu$_{3}$O$_{7-\delta}$ (YBCO) films \cite{AICH19,KARR24a}. Current studies are underway to extend this technique to thin Bi-2212 films, which must have a thickness of less than about 50\,nm to allow full penetration of the employed 30\,keV He$^+$ ions.

This study aims to provide valuable data on very thin Bi-2212 films by measurements of the anisotropic magnetoresistance with the magnetic field oriented perpendicular and parallel to the CuO$_2$ layers; the latter has been rarely explored so far. Critical current and Hall effect measurements provide additional insights into the transport properties of our samples. By presenting a detailed analysis of the electrical transport properties of Bi-2212 and providing estimates for the most important superconducting material parameters, we seek to contribute to the fundamental background to inspire future applications in electronic devices harnessing these remarkable materials.

\section*{Results and discussion}

\subsection*{Superconducting transition and magnetoresistance}

\begin{figure}[b!]
    \centering
    \includegraphics[width=0.85\textwidth]{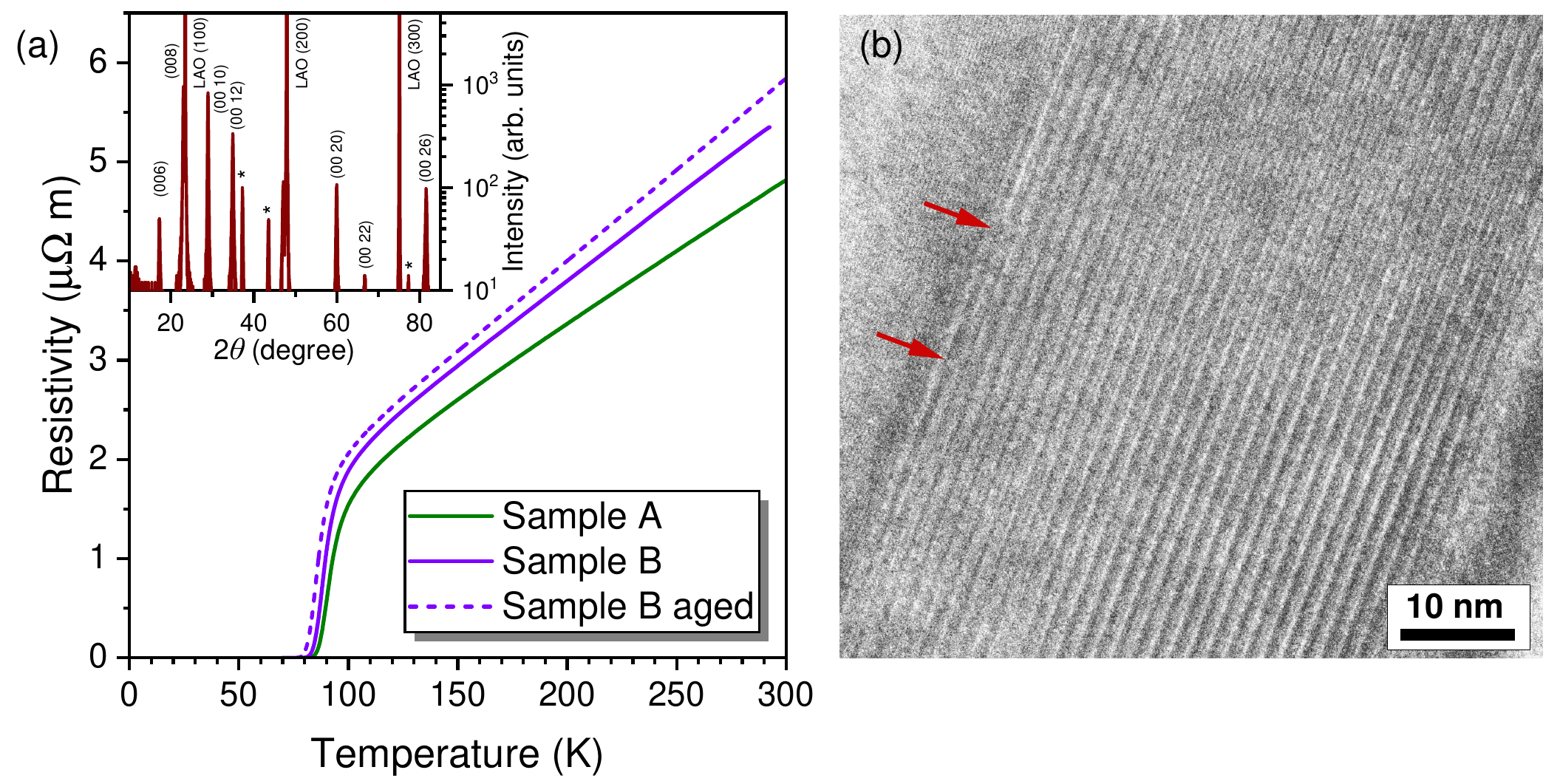}
    \caption{(\textbf{a}) Resistivity of the two Bi-2212 samples in zero magnetic field. The dashed line presents data of sample B collected 319 days later. Inset: X-ray diffraction $\theta-2\theta$ scans of another sample prepared identically. The asterisks mark peaks stemming from the aluminium sample holder. (\textbf{b}) Transmission electron micrograph of a 49-nm-thick Bi-2212 film revealing excellent layered structure. The stripe-like features correspond to half-unit cell periodicity. The grey area at the top left is the LAO substrate. The red arrows highlight some examples of growth defects at the substrate interface.}
    \label{fig:rho_TEM}
\end{figure}

The epitaxial growth of our Bi-2212 films was checked by X-ray diffraction $\theta-2\theta$ scans (see the inset of Fig.~\ref{fig:rho_TEM}a) on a film prepared with identical deposition parameters as sample A. The peaks could be allocated to the (0 0 $l$) reflections of Bi-2212, the  LaAlO$_3$ (LAO) substrate, and the Al sample holder while no other phases have been detected. This also demonstrates the good $c$-axis orientation of the unit cells grown on the crystalline substrate.

Cross-sectional transmission electron microscopy (TEM) analysis confirmed the excellent epitaxial growth and layered structure of our thin Bi-2212 films. The half unit-cell periodicity is clearly visible in Fig.~\ref{fig:rho_TEM}b, indicating that this 49-nm-thick Bi-2212 film consists of 16 unit cells stacked along the $c$ direction. The number of layers corresponds to a calculation using the crystallographic unit cell parameters and the thickness of the film as determined by atomic force microscopy (AFM). The TEM image reveals some structural defects at the interface of Bi-2212 film and LAO ($a_\mathrm{LAO} = 3.82$\,\r{A}) crystal substrate. Red arrows mark some examples. For comparison, TEM images of our Bi-2212 films with 13 unit cell thickness on (LaAlO$_3$)$_{0.3}$(Sr$_2$TaAlO$_6$)$_{0.7}$ (LSAT, $a_\mathrm{LSAT} = 3.87$\,\r{A}) substrates were very similar and showed the same layered structure of Bi-2212 and very similar nanoscopic defects at the film-substrate interface.

The resistivities $\rho_{xx}$ of the two samples are strictly linear from room temperature down to approximately 130\,K, as shown in Fig.~\ref{fig:rho_TEM}a. This behaviour is representative of optimally doped Bi-2212, whereas overdoped samples typically show a positive curvature and underdoped samples display a negative curvature in their $\rho_{xx}(T)$ curves \cite{KONS99a}. Below 130\,K, superconducting order parameter fluctuations introduce an additional paraconductivity contribution, causing $\rho_{xx}(T)$ to decrease as $T$ approaches  $T_c$ \cite{LARK05M}. The critical temperature $T_{c,inf}$, defined by the inflection point where $\mathrm{d}^2 \rho_{xx}/ \mathrm{d}T^2=0$ is $T_{c,inf}= 90.0$\,K (sample A) and $T_{c,inf}= 88.0$\,K (sample B). A linear extrapolation of the normal-state $\rho_{xx}(T)$ behaviour to $\rho_{xx}(T)=0$ yields slightly negative intercepts of about -30\,K (sample A) and -20\,K (sample B).

At room temperature, the resistivity $\rho_{xx}$(300\,K) closely matches that of optimally-doped Bi-2212 single crystals \cite{WATA97} and is only marginally higher than that reported for 200\,nm thick Bi-2212 films \cite{KONS99a}. This demonstrates that the substantial reduction in film thickness does not significantly increase the resistivity. Overall, these characteristics are consistent with those expected from high-quality, optimally doped copper-oxide superconductors.

Despite having a thickness of about 13 unit cells, our Bi-2212 films demonstrate remarkable stability. Following the initial measurement, we re-examined the temperature dependence of the resistivity of sample B after 319 days. This is illustrated as solid and dashed lines in {Fig.~\ref{fig:rho_TEM}a}, respectively. During the 319 days between measurements, the sample underwent several processes, including temperature cycling in both a vacuum and a helium atmosphere for various investigations. It was also mailed, rinsed in acetone, reconnected multiple times, and stored in a desiccator with dry air at room temperature. As a result of these procedures, the critical temperature experienced a drop of $\Delta T_\mathrm{c,inf} = 2.8\,\mathrm{K}$, and the normal state resistivity increased by 5\% at 150 K.

\begin{figure}[b!]
    \centering
    \includegraphics[width=0.9\textwidth]{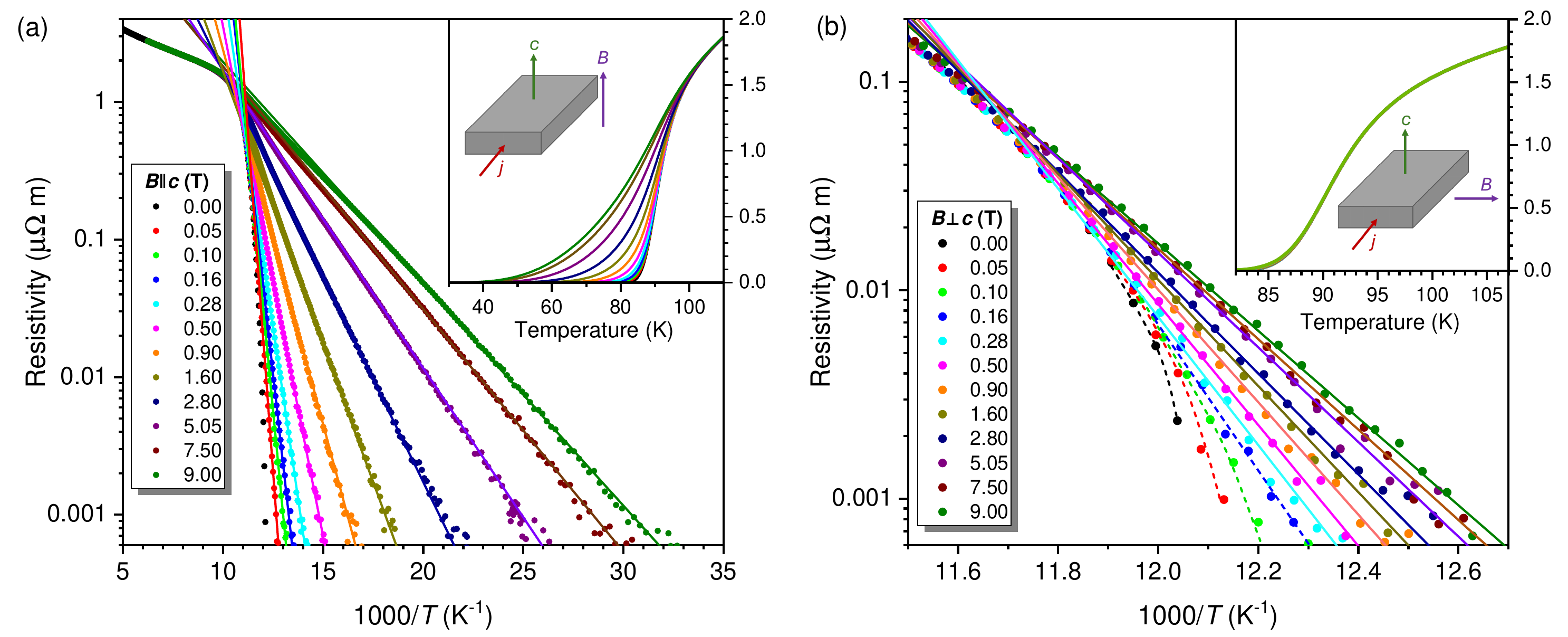}
    \caption{Arrhenius plots of the magnetoresistivity of sample A under various magnetic fields, oriented as shown in the schematics. The solid lines are fits to Eq.~\eqref{eq:TAFF}, illustrating the thermally activated flux-flow regime. The dashed lines are guides to the eye. The insets present the respective data on linear scales. (\textbf{a}) Fields applied perpendicular to the CuO$_2$ planes. Data adapted from \cite{KEPP23}. (\textbf{b}) Fields applied parallel to the CuO$_2$ planes. Note that all the curves nearly overlap in the inset.}
    \label{fig:MR}
\end{figure}

In external magnetic fields applied perpendicular to the film surface---and thus orthogonal to the CuO$_2$ layers---the superconducting transitions broaden significantly, forming the characteristic `fan-shape' commonly observed in copper-oxide superconductors. The inset of Fig.~\ref{fig:MR}a illustrates this pronounced effect, consistent with earlier results obtained on 400\,nm thick Bi-2212 films \cite{RI94}.

An Arrhenius plot of the magnetoresistivity, shown in Fig.~\ref{fig:MR}a, reveals linear behaviour for $T\lesssim 91$\,K, indicative of thermally-assisted flux flow (TAFF). In this regime, the resistivity can be described as
\begin{equation}
\rho_{\mathrm{TAFF}} \propto \exp \left(-U / k_{\mathrm{B}} T\right),
\label{eq:TAFF}
\end{equation}
where $U$ is the vortex pinning energy and $ k_{\mathrm{B}}$ the Boltzmann constant. Independent tests at selected points confirmed that $\rho_{xx}(B,T)$ is ohmic up to currents an order of magnitude larger than those used in the main measurements. In zero magnetic field, however, the Arrhenius law no longer applies, and the curve acquires a negative curvature, presumably due to the absence of Abrikosov vortices. Notably, in our film, the Arrhenius relation holds perfectly over three decades of $\rho_{xx}(B,T)$, closely resembling the behaviour reported in single-crystal experiments \cite{PALS88}. This contrasts earlier work on molecular-beam epitaxy (MBE) grown Bi-2212 films with 100\,nm thickness, which exhibited a slight negative curvature and thus introduced uncertainties in determining the pinning energy $U$ \cite{KUCE92}.

The magnetoresistivity is dramatically different when the magnetic field is oriented parallel to the CuO$_2$ planes, i.e., to the $ab$ direction. A broadening of the superconducting transitions is barely detectable in magnetic fields up to 9\,T (see the inset of Fig.~\ref{fig:MR}b). The Arrhenius law of Eq.~\eqref{eq:TAFF} can be observed only for $\mu_0H\gtrsim 0.3$\,T while a negative curvature prevails at smaller magnetic fields, indicated by dashed lines in Fig.~\ref{fig:MR}b.

\begin{figure}[t!]
    \includegraphics[width=0.5\columnwidth]{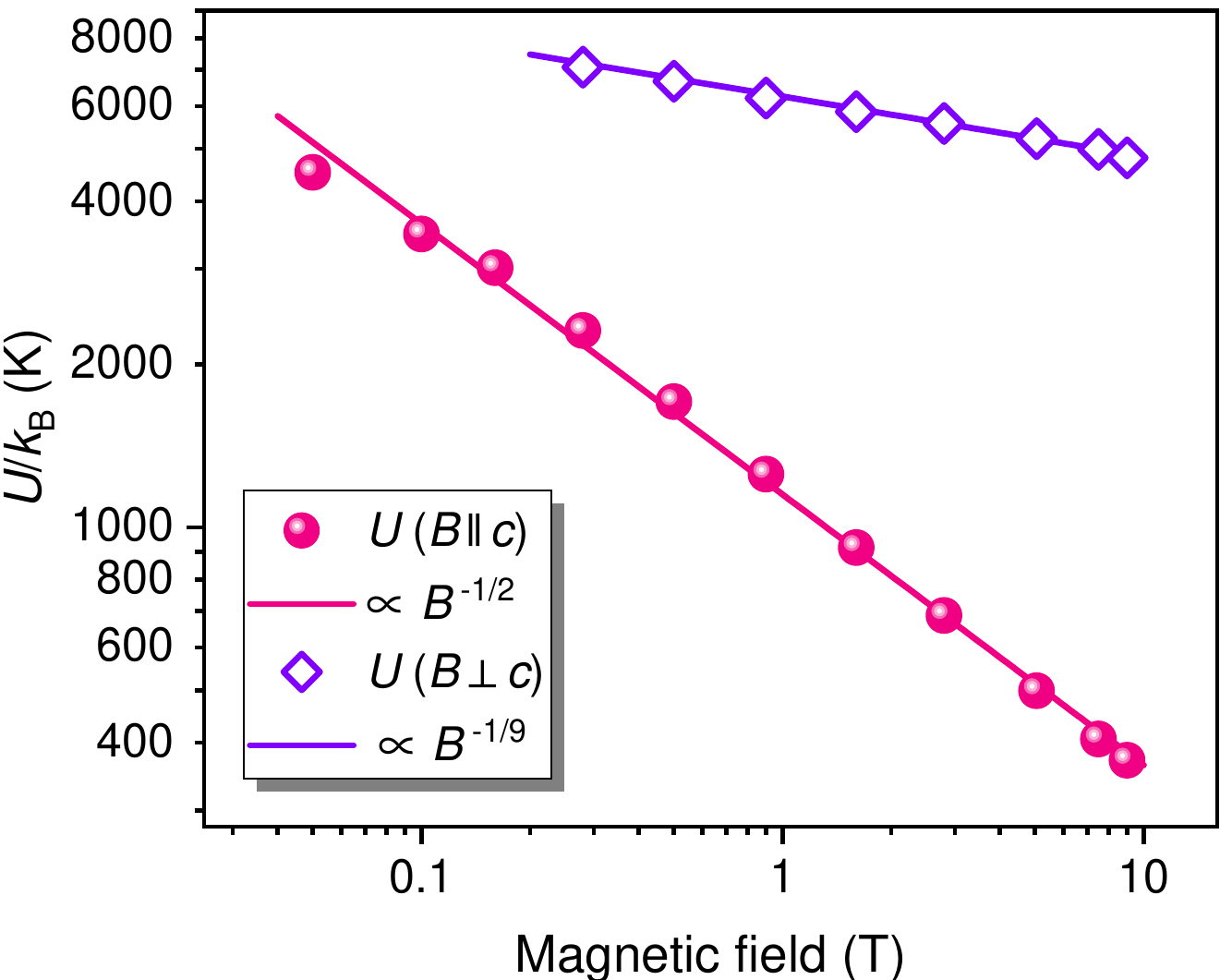}
    \caption{Vortex pinning energy $U/k_{\mathrm{B}}$, expressed in temperature units, determined from the fits to Eq.~\eqref{eq:TAFF} using the data displayed in Fig.~\ref{fig:MR} (sample A). The lines represent power-law fits to the resulting $U(B)$ dependencies.}
    \label{fig:U}
\end{figure}

Furthermore, the pinning energy $U$ shows a pronounced anisotropy, as illustrated in Fig.~\ref{fig:U}. Note that, due to the experimental geometry, a Lorentz force on vortices acts in both orientations of $B$, as can be seen from the schematics in {Fig.~\ref{fig:MR}}. For $B \parallel c$, $U$ is roughly twice as large as in Bi-2212 single crystals \cite{PALS88,BUSC92} and follows a $B^{-1/2}$ dependence, suggesting that it is governed by the average lateral spacing of pancake vortices. This behaviour differs markedly from the $U \propto 1/B$ trend commonly observed in thin films of less-anisotropic REBa$_{2}$Cu$_{3}$O$_{7-\delta}$ (RE-123, RE\,=\,rare earth element) superconductors \cite{SUN89}, thereby underscoring the distinct vortex dynamics associated with Josephson-coupled pancake vortices in Bi-2212 compared to conventional Abrikosov vortices in RE-123.

For $B \parallel ab$, $U$ is notably higher and exhibits only a weak dependence on $B$, as illustrated by the nonphysical relationship $B^{-1/9}$. This peculiar behaviour presumably arises from the strong intrinsic pinning of Josephson vortices (JVs) by the stacked CuO$_2$ layers, effectively preventing JV hopping along the crystallographic $c$ axis. Remarkably, at 1\,T, the pinning energy in our film is almost an order of magnitude greater than that reported for Bi-2212 single crystals \cite{PALS88}, and it decreases far more slowly as $B$ increases. One possible explanation for the observed difference is related to the significantly greater thickness of single crystals compared to our very thin films. In the orientation $B \parallel ab$, the magnetic field may not penetrate uniformly into single crystals. In contrast, this penetration is much more effective in thin films, ensuring that the pinning potential in each layer is fully active.

The $\rho_{xx}(T,B)$ data of sample A, shown in Fig.~\ref{fig:MR}, can be used to estimate the zero-temperature upper critical fields $B_\mathrm{c2}^{\perp}(0)$ and $B_\mathrm{c2}^{\parallel}(0)$ for $B$ perpendicular and parallel to the CuO$_2$ layers, respectively. To this end, we define $B_\mathrm{c2}(T)$ as the field at which $\rho_{xx}(T)=0.5\, \rho_{xx}(120\,\mathrm{K})$. Subsequently, the Helfand-Werthamer relation for moderately dirty superconductors \cite{HELF66} is employed to extrapolate these critical fields to $T = 0\, \mathrm{K}$.
\begin{equation}
\left.B_{\mathrm{c} 2}(0) \approx 0.7\, T_{\mathrm{c}} \frac{\mathrm{d} B_{\mathrm{c} 2}}{\mathrm{~d} T}\right|_{T_{\mathrm{c}}}; B_{\mathrm{c} 2}(T) \geq 1\,\mathrm{T}.
\end{equation}
The results for sample A are $B_\mathrm{c2}^{\perp}(0)\approx 106$\,T and $B_\mathrm{c2}^{\parallel}(0) \approx 19,000$\,T, revealing an anisotropy parameter $\gamma=B_\mathrm{c2}^{\parallel}(0)/B_\mathrm{c2}^{\perp}(0) \approx 180$. For sample B we find $B_\mathrm{c2}^{\perp}(0)\approx 86$\,T.

Using the Ginzburg-Landau relations
\begin{equation}
B_{\mathrm{c} 2}^{\perp}(0)=\frac{\Phi_0}{2 \pi \xi_{ab}^2} \quad \mathrm{and} \quad B_{\mathrm{c} 2}^{\parallel}(0)=\frac{\Phi_0}{2 \pi \xi_{ab}\xi_{c}}
\end{equation}
we obtain the zero-temperature Ginzburg-Landau coherence lengths $\xi_{ab} \approx 1.8$\,nm and $\xi_{c} \approx 0.01$\,nm in the in-plane and out-of-plane directions for sample A, respectively, and $\xi_{ab} \approx 2.0$\,nm for sample B. The extremely small $\xi_{c}$ reflects the large anisotropy of Bi-2212 and is notably smaller than even in oxygen-depleted YBa$_{2}$Cu$_{3}$O$_{6.6}$ \cite{GOB95a}.

An alternative method to estimate coherence lengths involves analysing the paraconductivity and magnetoconductivity induced by superconducting order parameter fluctuations. Such studies in single crystals and thin films have reported  $\xi_{ab}$ values ranging from 0.9\,nm to 3.8\,nm and $\xi_{c}$ between $\lesssim 0.04$\,nm and 0.1\,nm \cite{POMA96}. Our even smaller estimate for $\xi_{c}$ confirms the excellent CuO$_2$-layer stacking in our films and can be attributed to the careful alignment of the magnetic field with these layers.

\subsection*{Critical current}

\begin{figure}[t!]
    \centering
    \includegraphics[width=0.9\textwidth]{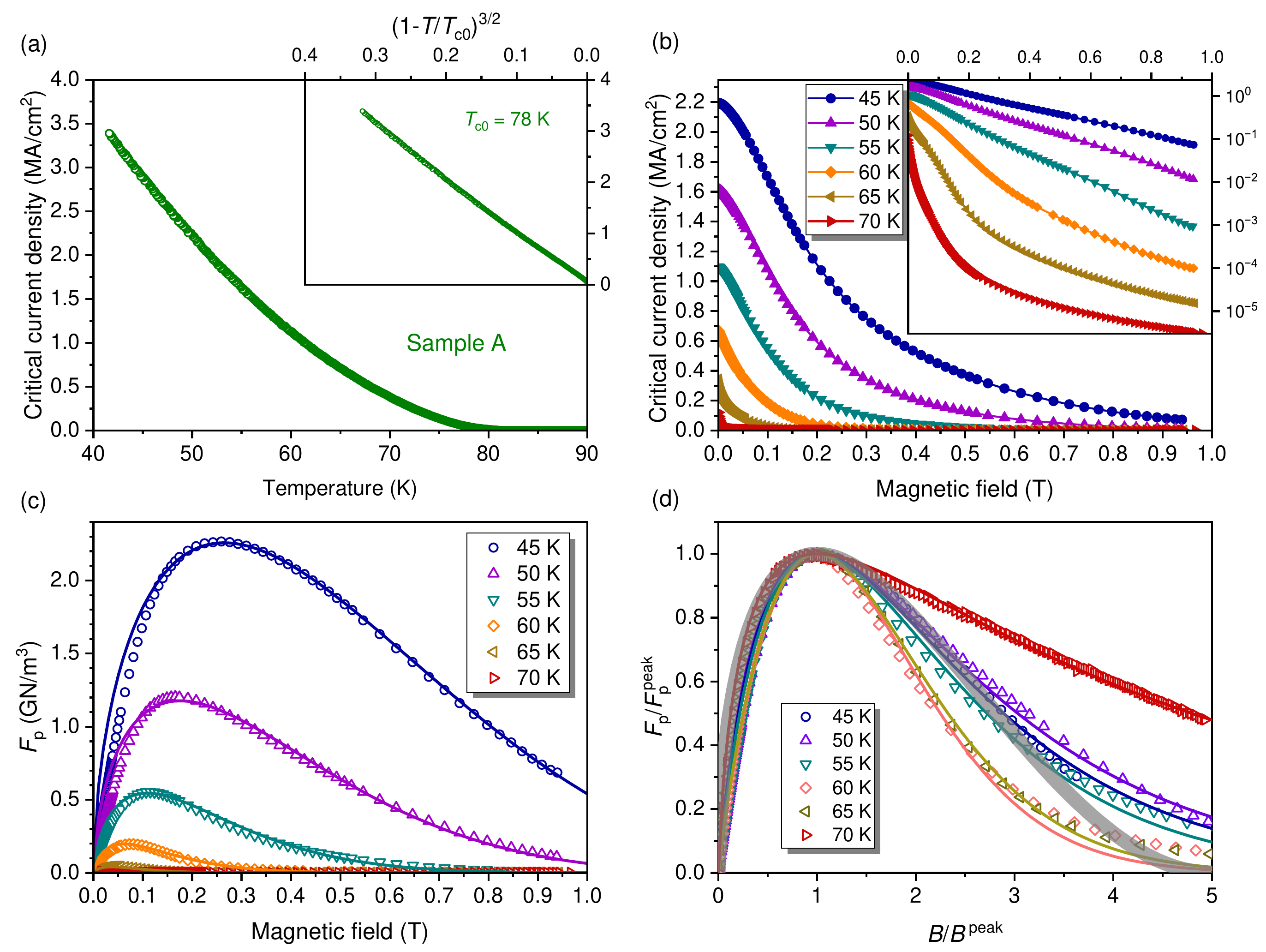}
    \caption{(\textbf{a}) Critical current density of sample A under self-field conditions. Inset: Same data plotted as a function of reduced temperature. The resulting linear relationship confirms the validity of Eq.~\eqref{eq:JcGL}. (\textbf{b}) Critical current density of sample B as a function of the external magnetic field at various temperatures. Inset: Semi-logarithmic plot of the data in the main panel. (\textbf{c}) Volume vortex pinning force of sample B as a function of the magnetic field at various temperatures. The solid lines represent fits to the power-law model as given by Eq.~\eqref{eq:KDH}. (\textbf{d}) Normalized volume vortex pinning force of sample B. The solid lines represent fits to the exponential decay model of Eq.~\eqref{eq:exp} with a fixed $p=0.9$ and $\varepsilon$ as plotted in Fig.~\ref{fig:KDH}d. For comparison, the broad grey line illustrates the behaviour predicted by the power-law model [Eq.~\eqref{eq:KDH}] with the parameters $p=0.5$ and $q=2$.}
    \label{fig:jc}
\end{figure}

In the following, we consider thin films whose thickness $t_z$ is much smaller than both their lateral dimensions and the zero-temperature London penetration depths along the $ab$ plane, $\lambda_{ab}$, and perpendicular, $\lambda_{c}$. Under these conditions, the superconducting current density $J$ can be assumed to be homogeneous within the film. Fig.~\ref{fig:jc}a presents the critical current density $J_\mathrm{c}^{\mathrm{sf}}$ in the absence of an external magnetic field, i.e., in the current's self-field $B=\mu_0 J t_z/2$ \cite{BABA05}. The inset shows that the data conform well to the relation
\begin{equation}
J_{\mathrm{c}}^{\mathrm{sf}}(T)=J_\mathrm{c}^{\mathrm{sf}}(0)\left(1-\frac{T}{T_\mathrm{c0}}\right)^{3/2},
\label{eq:JcGL}
\end{equation}
where $J_\mathrm{c}^{\mathrm{sf}}(0)$ is the critical current density at zero temperature and $T_\mathrm{c0}$ is a fit parameter associated with the temperature at which $\rho_{xx}$ vanishes [see Fig.~\ref{fig:rho_TEM}a]. Although Eq.~\eqref{eq:JcGL} strictly applies only near $T_\mathrm{c0}$ \cite{KUPR80}, it provides a good description of our data down to 42\,K for sample A. However, deviations from this trend occur towards lower temperatures, precluding the determination of $J_\mathrm{c}^{\mathrm{sf}}(0)$ from Eq.~\eqref{eq:JcGL}.

A sophisticated model for thin-film microbridges has been proposed and described in detail elsewhere \cite{TALA15,TALA17}, where $J_{\mathrm{c}}^{\mathrm{sf}}$ is determined by the lower critical field as $J_{\mathrm{c}}^{\mathrm{sf}}=H_{\mathrm{c1}}^{\perp} / \lambda_{ab}$, leading to an expression confirmed in many superconducting materials
\begin{equation}
J_\mathrm{c}^\mathrm{sf}(0)=\frac{\Phi_0}{4 \pi \mu_0 \lambda_{ab}^3}\left(\ln \frac{\lambda_{ab}}{\xi_{ab}} + 0.5\right).
\label{eq:JcTALA}
\end{equation}
With a computer code \cite{CRUM16} based on the afore-mentioned model for fitting $J_{\mathrm{c}}^{\mathrm{sf}}(T)$ data and considering the $d$-wave gap symmetry in our samples, the extrapolation to $T=0$ yields $J_\mathrm{c}^\mathrm{sf}(0) \approx 8.1\,\mathrm{MA/cm}^2$. Of course, an accompanying estimate of the minimum $J_\mathrm{c}^\mathrm{sf}(0)$ in our two Bi-2212 films is also warranted. For this lower bound, low-temperature measurements revealed $J_\mathrm{c}^\mathrm{sf}(6\,\mathrm{K}) = 5.1\,\mathrm{MA/cm}^2$ in sample B, though this value is subject to some uncertainty due to heating from dissipation at the current contacts. Even so, any systematic temperature error would lead to an underestimate of $J_{\mathrm{sf}}(0)$, thereby reinforcing our conservative lower-limit estimate of $J_\mathrm{c}^\mathrm{sf}(0) \ge 5.1\,\mathrm{MA/cm}^2$.

By inserting the previously determined values $\xi_{ab} = 1.8$\,nm for sample A and $\xi_{ab} = 2.0$\,nm for sample B into Eq.~\eqref{eq:JcTALA}, we obtain an estimate for $\lambda_{ab} \sim 200$\,nm (sample A) and $\lambda_{ab} \lesssim 240$\,nm (sample B). Furthermore, applying the mean-field temperature dependence $\lambda_{ab}(T) \propto \left(1-T / T_c\right)^{-1 / 2}$ in Eq.~\eqref{eq:JcTALA} yields the temperature dependence shown in the inset of Fig.~\ref{fig:jc}a.

Now we focus on the critical current density of Bi-2212 films in a magnetic field applied parallel to the $c$ axis. As previously observed \cite{SCHM90,LABD92,ZHAO24}, $J_\mathrm{c}(B,T)$ decreases rapidly with increasing field strength. This trend is evident in Fig.~\ref{fig:jc}b, where the main panel presents $J_c(B)$ on a linear scale, and the inset shows the same data on a semi-logarithmic scale. For instance, at $45\,\mathrm{K} \sim T_\mathrm{c}/2$, $J_\mathrm{c}$ diminishes approximately one order of magnitude up to 1\,T, following an almost exponential decay.

A direct comparison with data on other Bi-2212 films \cite{LABD92} can be made at 50\,K. Our 42-nm film exhibits  $J_{\mathrm{c}}^{\mathrm{sf}}=1.6\,\mathrm{MA/cm}^2$ and a ratio $J_{\mathrm{c}}^{\mathrm{sf}}/J_{\mathrm{c}}(1\,\mathrm{T}) \sim 100$, while a 100\,nm film \cite{LABD92} shows $J_{\mathrm{c}}^{\mathrm{sf}}\simeq 0.06\,\mathrm{MA/cm}^2$ and $J_{\mathrm{c}}^{\mathrm{sf}}/J_{\mathrm{c}}(1\,\mathrm{T}) \sim 200$. Additionally, another study on a 50-nm film \cite{SCHM90} reported $J_{\mathrm{c}}^{\mathrm{sf}}=1.0\,\mathrm{MA/cm}^2$, about half of our sample A's value of $J_{\mathrm{c}}^{\mathrm{sf}}(50\,\mathrm{K})=2.2\,\mathrm{MA/cm}^2$.

Surprisingly, Bi-2212 films containing self-assembled NiO nanorods exhibit a $J_\mathrm{c}^{\mathrm{sf}}$ at $\sim T_\mathrm{c}/2$ that is about two orders of magnitude lower than that of our samples. Nonetheless, despite the presence of these artificial pinning centres, the relative field-dependent decay $J_\mathrm{c}(B)$ is very similar to our samples \cite{ZHAO24}. This suggests that our very thin Bi-2212 films provide significantly stronger intrinsic pinning than typically observed in single crystals \cite{BERG89} or even in most 100-nm thick films \cite{LABD92}. It is well-established that nanostrain can enhance vortex pinning in HTS \cite{LI19}. In the case of these very thin films, the lattice mismatch with the LAO substrate induces a uniform misfit strain that relaxes by forming structural defects, causing localized strain. However, complete relaxation of misfit strain occurs only at some distance from the interface, which makes a higher density of defects more likely in thinner films. As a result, this could result in higher $J_\mathrm{c}$ values compared to thicker films.

The $J_{\mathrm{c}}(B,T)$ data can be further analysed in terms of the volume pinning force $F_\mathrm{p}(B,T)$ by a well-established, semi-phenomenological scaling approach \cite{KRAM73,DEWH74} that is represented by the Kramer–Dew–Hughes model
\begin{equation}
F_\mathrm{p}(B,T) = J_{\mathrm{c}}(B,T) B = F_\mathrm{p}^* \left(\frac{B}{B^*}\right)^p\left(1-\frac{B}{B^*}\right)^q.
\label{eq:KDH}
\end{equation}
In metallic superconductors, $B^*$ corresponds to $B_\mathrm{c2}$. For HTS, however, it is more appropriately associated with the irreversibility field $B_\mathrm{irr}$. $F_\mathrm{p}^*$ is a temperature-dependent scaling parameter, while the dimensionless exponents $p$ and $q$ reflect the specific nature and dimensionality of the pinning centres---whether they are point-like, linear, or planar defects.

In a simplified model, each vortex experiences a pinning force per unit length $f_\mathrm{p}$. At moderate fields, where vortices are sufficiently dilute, the volume pinning force can be approximated by $F_\mathrm{p}=\rho_A f_\mathrm{p}$, where $\rho_A$ is the areal density of vortices. Since the inter-vortex spacing scales as $B^{1/2}$ and the number of vortices within a given volume scales with $B$, theoretical considerations suggest that $p$ is in the range $0.5 \dots 1.0$ \cite{DEWH74}. At higher magnetic fields, collective vortex effects become more significant. It has been argued that, under these conditions, the second factor in Eq.~\eqref{eq:KDH} is dominated by the softening of the shear modulus $c_{66}$ of the flux line lattice \cite{KRAM73}, which leads to a value $q=2$.

Fig.~\ref{fig:jc}c shows the field dependence of the volume pinning force $F_\mathrm{p}(B,T)$, which exhibits the characteristic shape commonly observed in HTS. In comparison to YBCO, the maximum pinning force  $F_\mathrm{p}^\mathrm{max}$ in Bi-2212 thin films is notably smaller and occurs at approximately one-tenth the magnetic field $B$. This shift and reduction are not unexpected, given the substantially weaker pinning of two-dimensional, pancake-like vortices in Bi-2212 and the absence of strong, micro-twin-boundary-induced pinning centres that are known to enhance $F_\mathrm{p}(B,T)$ in YBCO.

\begin{figure}[b!]
    \includegraphics[width=0.6\columnwidth]{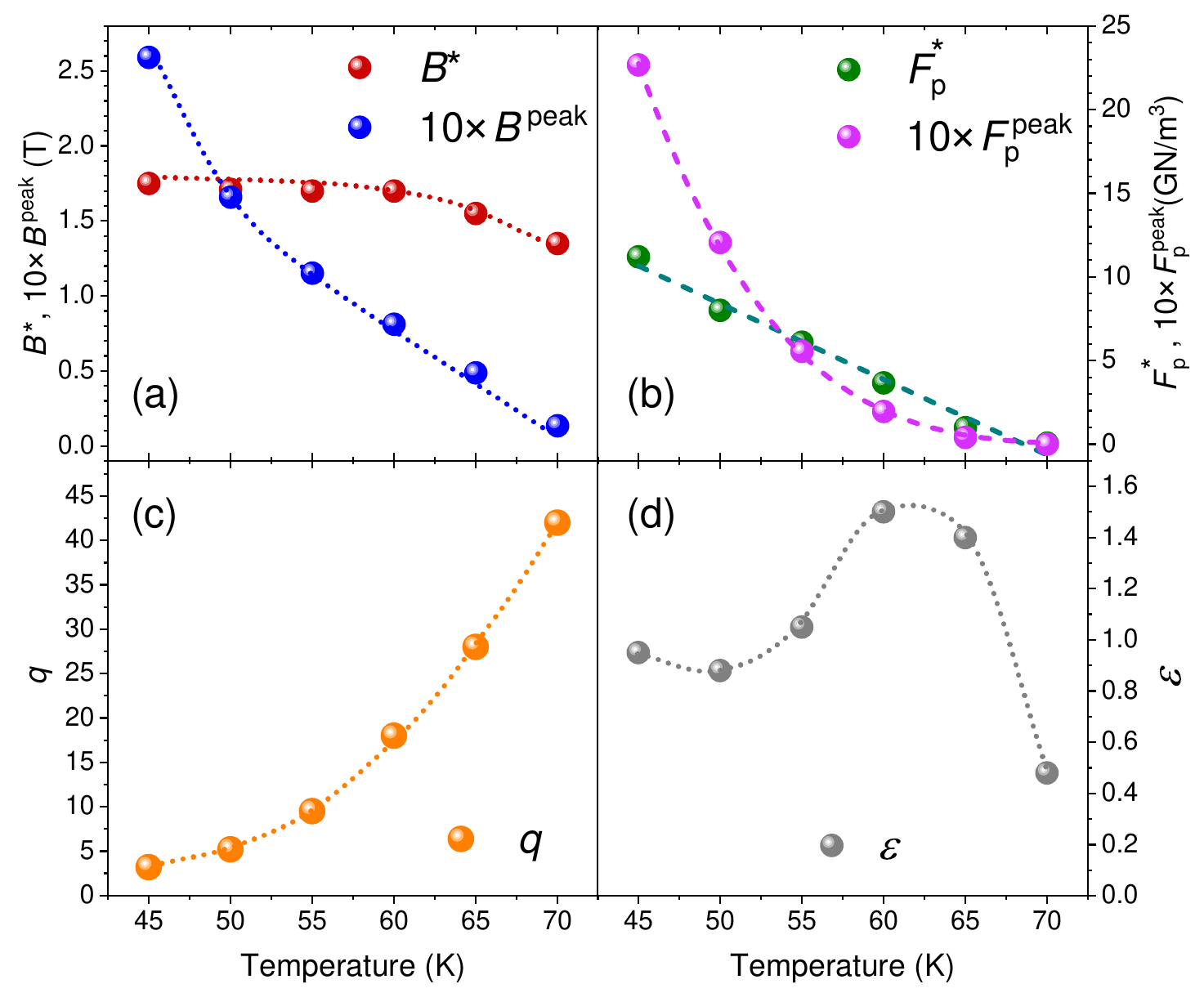}
    \caption{Fit parameters obtained from the power-law and exponential decay models. (\textbf{a}) Magnetic field $B^\mathrm{peak}$ corresponding to the maximum in the volume pinning force, along with the magnetic field scaling parameter $B^*$. (\textbf{b}) Peak value of the volume vortex pinning force $F_\mathrm{p}^\mathrm{peak}$ and its associated scaling parameter $F_\mathrm{p}^*$. (\textbf{c}) Parameter $q$ of Eq.~\eqref{eq:KDH}. (\textbf{d}) Parameter $\varepsilon$ of Eq.~\eqref{eq:exp}. The dashed lines represent a linear fit for $F_\mathrm{p}^*$ and a power-law fit for $F_\mathrm{p}^*$. Dotted curves are guides to the eye.}
    \label{fig:KDH}
\end{figure}

The temperature dependencies of the fit parameters $B^*$, $F_\mathrm{p}^*$, and $q$ are shown in Fig.~\ref{fig:KDH}. While the magnetic field associated with maximum pinning force, $B^\mathrm{peak}$, decreases substantially as the temperature increases, the scaling parameter $B^*$ exhibits only a minor dependence, as shown in Fig.~\ref{fig:KDH}a. A similar pattern emerges in Fig.~\ref{fig:KDH}b for $F_\mathrm{p}^\mathrm{peak}$ accompanied by a much weaker temperature dependence of $F_\mathrm{p}^*$. The parameter $p = 0.57 \dots 0.68$ remains nearly constant with respect to temperature, indicating that point-like defects dominate the pinning mechanism. In contrast, $q$ exhibits an unphysical deviation from the theoretically expected value of $q \sim 2$ once $T > 45\,\mathrm{K}$, as displayed in Fig.~\ref{fig:KDH}c. This anomaly, combined with the unexpectedly weak temperature dependence of $B^*$, which should closely follow the irreversibility field  $B_\mathrm{irr}$, suggests that the model in Eq.~\eqref{eq:KDH} may not be fully applicable to Bi-2212 thin films, despite the seemingly good fits presented in Fig.~\ref{fig:jc}(c).

The exponential decrease of resistivity with inverse temperature, as depicted in Fig.~\ref{fig:MR}a, strongly suggests that an exponential decay model \cite{FABB96,JIRS00} provides a more accurate representation of the observed data. By normalizing the data to the experimentally determined values $F_\mathrm{p}^\mathrm{peak}$ and $B^\mathrm{peak}$, the model effectively reduces to just two adjustable parameters, $p$ and $\varepsilon$. The latter is connected to the slope of $J_\mathrm{c}$(B), as discussed elsewhere {\cite{JIRS00}}.
\begin{equation}
\frac{F_\mathrm{p}}{F_\mathrm{p}^\mathrm{peak}} = \left( \frac{B}{B^\mathrm{peak}} \right)^p \exp \left\{ \left[ 1- \left( \frac{B}{B^\mathrm{peak}} \right)^\varepsilon \right] \frac{p}{\varepsilon} \right\}.
\label{eq:exp}
\end{equation}

 As demonstrated in Fig.~\ref{fig:jc}d, employing a fixed $p = 0.9$ and a parameter range $0.5 \leq \varepsilon \leq 1.5$ (see Fig.~\ref{fig:KDH}d) yields satisfactory agreement with the experimental data. This normalized representation also facilitates a more robust assessment of the fit quality across a wide range of temperatures. For comparison, the broad grey line shows the prediction of the power-law model [Eq.~\eqref{eq:KDH}] with the parameters $p=0.5$ and $q=2$. Although the power-law model reproduces experimental results well up to $B^\mathrm{peak}$, qualitative inspections indicate that it cannot capture the extended tails of $F_\mathrm{p}$ at $B \gg B^\mathrm{peak}$. A similar shortcoming was also observed in Bi-2223 thin films \cite{YAMA93}. The inability of the power-law model to accurately describe the volume pinning force in Bi-2212 highlights the distinctive vortex dynamics in this material. Compared to YBCO and other less anisotropic superconductors, where power-law scaling has been successfully applied \cite{TALA22}, Bi-2212 exhibits more complex behaviour.

\subsection*{Hall effect}

\begin{figure}[b!]
  \centering
  \includegraphics[width=\textwidth]{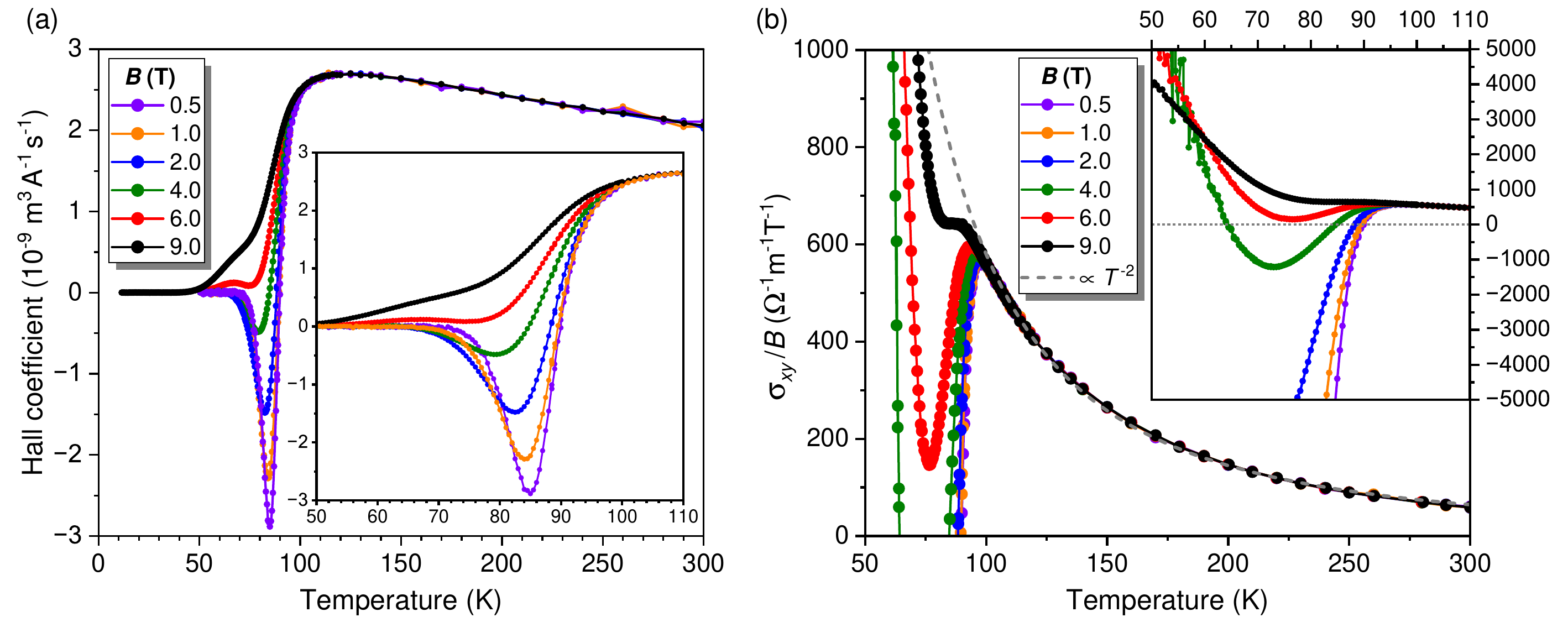}
  \caption{%
    (\textbf{a}) Hall coefficient measured in multiple magnetic fields applied perpendicular to the $ab$ planes of the Bi-2212 thin film (sample B). The inset provides a magnified view to emphasize the sign reversal of the Hall effect. (\textbf{b}) Normalized Hall conductivity $\sigma_{xy}/B$ as a function of temperature under various magnetic fields. The grey dashed line illustrates a $T^{-2}$ dependence of the normal-state $\sigma_{xy}/B$. Inset: Detailed view of the Hall effect sign change.}
  \label{fig:hall}
\end{figure}

One of the most intriguing phenomena observed in HTS is their unconventional Hall effect, closely associated with the `strange-metal' state observed predominantly in optimally doped HTS \cite{BARI22}. In our Bi-2212 thin films, this unconventional behaviour is even more pronounced. Unlike conventional metals, where the Hall coefficient remains relatively constant with temperature and serves as a direct measure of carrier type and density, Fig.~\ref{fig:hall}a demonstrates that the situation is considerably more complex in Bi-2212. 

The Hall coefficient is defined as $R_\mathrm{H} = E_y/(j_x B_z)$, where $E_y$ is the transverse electric field measured between the opposite side arms of the bridge, $j_x$ the current density along the bridge, and $B_z$ the magnetic field applied perpendicular to the film's surface. Data were collected on a rectangular-shaped bridge (sample B) with length $l_x=657\,\mu$m, width $w_y=102\,\mu$m, and thickness $t_z=42$\,nm. The positive sign of $R_\mathrm{H}$ confirms hole-like carriers as the dominant charge carriers. As the temperature decreases, $R_\mathrm{H}$ initially increases until a pronounced negative curvature sets in around 120\,K, ultimately giving way to a sharp drop at $T_\mathrm{c}(B)$. These features can be attributed to a negative contribution from superconducting order-parameter fluctuations \cite{PUIC04}, underscoring the intricate interplay between the normal-state and superconducting regimes in these materials.

\begin{figure}[t!]
  \includegraphics[width=0.5\columnwidth]{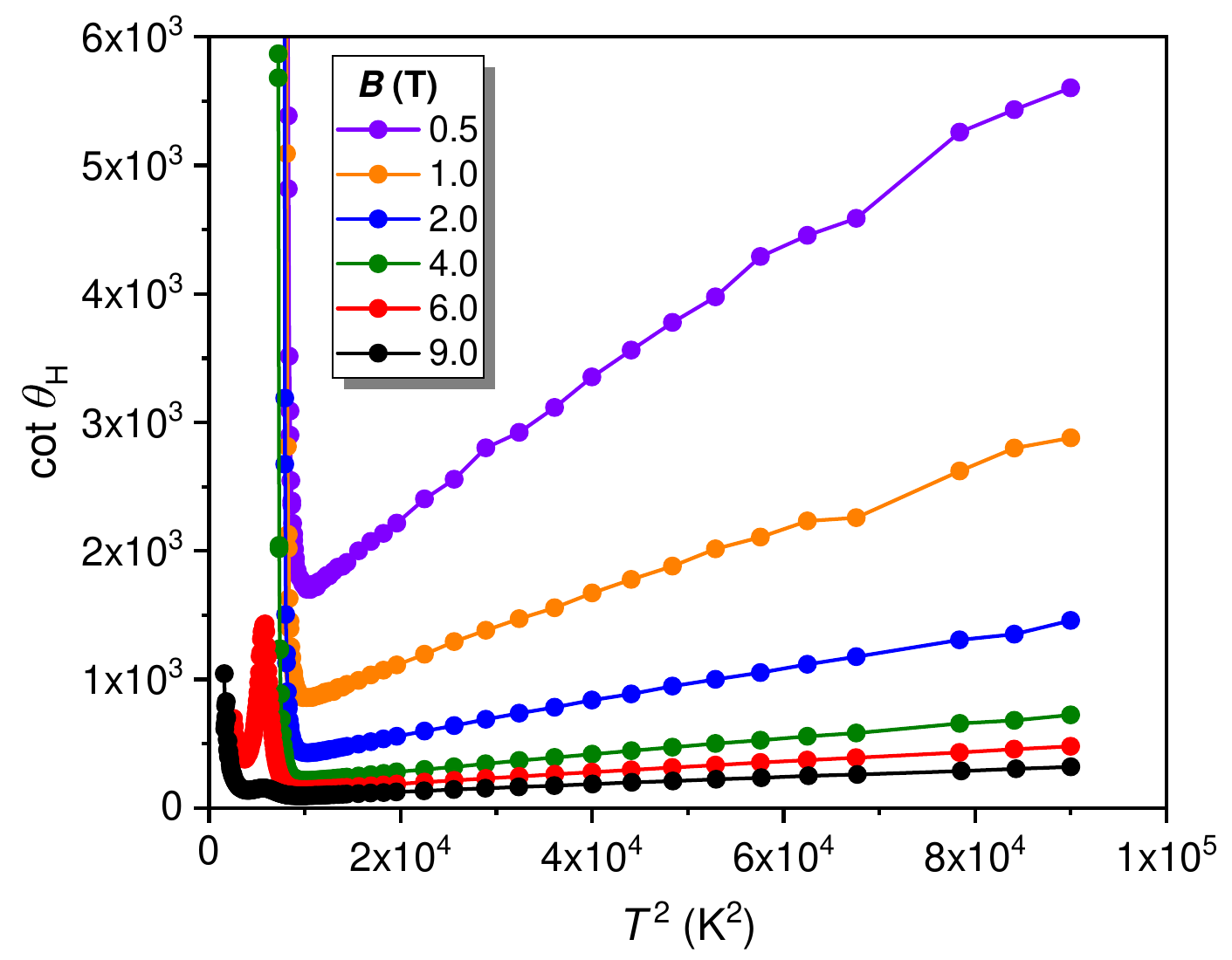}
  \caption{%
   Reciprocal tangent of the Hall angle plotted as a function of the squared temperature (Sample B).}
  \label{fig:cot}
\end{figure}

The carrier mobility inferred from the Hall effect $\mu_\textrm{H} = \tan \theta_\mathrm{H}/B_z = R_\mathrm{H}/\rho_{xx}$, where $\tan \theta_\mathrm{H} = E_y/E_x$ is the Hall angle, decreases with increasing temperature. A characteristic yet unconventional trend commonly observed in HTS is $\cot \theta_\mathrm{H} = \alpha T^2 +C$, where $C$ is proportional to the density of carrier-scattering defects, and $\alpha$ relates to the inverse carrier density \cite{ANDE91}. As shown in Fig.~\ref{fig:cot}, this relationship also holds for our Bi-2212 films. Extrapolation of the data to $T^2=0$ gives $\cot \theta_\mathrm{H}(1\,T,0\,\mathrm{K}) = 0.06$ whereas in 100\,nm to 200\,nm-thick films $\cot \theta_\mathrm{H}(1\,\mathrm{T},0\,\mathrm{K}) = 0.01$ has been reported \cite{KONS00}. We attribute our higher value to an increased density of scattering defects, possibly arising from lattice mismatch at the substrate interface.
Such defects, which cannot fully relax in ultra-thin films, may create stacking faults that lead to stronger vortex pinning and thus explain the exceptionally high critical currents observed in our samples.

While $R_\mathrm{H}$ in the normal state is field-independent (at least up to 9\,T), it becomes strongly $B$-dependent near and below the superconducting transition. Most remarkably, a sign change of $R_\mathrm{H}$ occurs in the vortex-liquid regime as displayed in Fig.~\ref{fig:hall}a. It was initially reported in YBCO and BSCCO ceramics \cite{GALF89} and is now recognized as a general feature of HTS, irrespective of their morphology. Various theories have been proposed to explain this Hall anomaly, including renormalized superconducting fluctuations \cite{PUIC04}, collective vortex effects \cite{AO98}, and the role of pinning centres \cite{KOPN99}. Still, a consensus has not yet emerged \cite{ZHAO19,GUO22}.

The amplitude of the negative $R_\mathrm{H}$ peak decreases markedly with increasing $B_z$, and by $B_z \geq 6$\,T, the sign change disappears. Still, a minor dip remains visible in the inset of Fig.~\ref{fig:hall}a. Most remarkably, at $B_z = 0.5$\,T, the negative Hall signal surpasses even the maximum positive $R_\mathrm{H}$ observed in the normal state. This effect is far more pronounced in our very thin Bi-2212 films than reported elsewhere, probably challenging theoretical models of the Hall anomaly based on the normal-state Hall effect in the cores of vortices. Our findings may encourage further attempts to distinguish between the proposed mechanisms for the Hall anomaly.

To properly interpret these results, it is important to consider the magneto-conductivity tensor. The measured quantities $\rho_{xx}$ and $\rho_{yx} = R_\mathrm{H} B_z$ relate to their conductivity counterparts via tensor inversion: $\rho_{xx} = \sigma_{xx}/(\sigma^2_{xx}+\sigma^2_{xy})$ and $\rho_{yx} = \sigma_{xy}/(\sigma^2_{xx}+\sigma^2_{xy})$. In the materials and magnetic fields under investigation $\sigma_{xx} \gg \sigma_{xy}$. Thus, $\rho_{xx} \simeq \sigma^{-1}_{xx}$ and
$\rho_{yx} \simeq \sigma_{xy}/\sigma^2_{xx}$. Even if $\sigma_{xy}$ in the vortex-liquid regime is believed to be largely unaffected by pinning \cite{VINO93}, the observable  $R_\mathrm{H}$ depends strongly on it \cite{PUIC04}.

It is illustrative to discuss a normalized quantity, $\sigma_{xy}/B$, presented in Fig.~\ref{fig:hall}b. In the normal state, $\sigma_{xy}/B \propto T^{-2}$. Just above  $T_\mathrm{c}$ a sharp negative trend emerges, producing dramatically larger negative values of $\sigma_{xy}/B$ than any positive values observed in the normal state for $B \leq 2$\,T [see the inset in Fig.~\ref{fig:hall}b]. A possible explanation of this effect and a quantitative model of the shape of the respective $R_\mathrm{H}(B,T)$ curves has been presented elsewhere \cite{PUIC04}. The model is based on superconducting order-parameter fluctuations and a low-temperature cutoff by a diverging $\sigma_{xx}$ due to pinning.

However, in Bi-2212 another twist occurs: at 4\,T, we observe a second sign reversal back to positive $\sigma_{xy}/B$. The effect may also exist in a broader $B$ range, but at lower fields, the Hall signal falls below the measurement resolution, and at higher fields, a residual dip, too weak to induce a sign reversal, is still detectable. Such a double sign change is notably pronounced in few-unit-cell Bi-2212 whiskers \cite{ZHAO19}, and we also see it in our approximately 13-unit-cell-thick films at similar temperatures.

Some observations in YBCO thin films hint that this second sign change may be related to pinning. In low fields, where intrinsic pinning is strong, a second sign reversal emerges in YBCO \cite{GOB00}. Furthermore, YBCO samples patterned with a periodic pinning array via masked ion irradiation also exhibit a second Hall sign reversal \cite{ZECH18a}. More studies are required to clarify the origin of this phenomenon in Bi-2212, and very thin films, as reported here, could be an excellent material for this endeavour.

Finally, Table~\ref{tab:params} summarizes the parameters obtained from our various measurements. Despite apparent differences between fabrication of the two samples, the fundamental superconducting parameters, including the upper critical field $B_\mathrm{c2}^{ab}(0)$, in-plane coherence length $\xi_{ab}$, London penetration depth $\lambda_\mathrm{L}^{ab}$, and the Ginzburg-Landau parameter $\kappa$, differ by no more than about 30\,\%. This consistency underscores the reproducibility of our results and provides a solid foundation for future studies on the properties of few-unit-cell thin Bi-2212 films.

\section*{Methods}

\subsection*{Fabrication of Bi$_2$Sr$_2$CaCu$_2$O$_{8+x}$ thin films}

Pulsed-laser deposition (PLD) is the preferred method to fabricate high-quality thin films of Bi-2212 as described in more detail elsewhere \cite{KEPP23}. Briefly, a substrate temperature of $T = 780\,^\circ$C and  oxygen background pressure of $p(\text{O}_2) = 1.2$\,mbar was used. Post-deposition \emph{in-situ} annealing of the films at $T = 600\,^\circ$C ensured optimally-doped samples. Single-crystalline (0~0~1) LaAlO$_3$  substrates with a lattice parameter $a_L = 3.82$\,\AA\ were used. The number of laser pulses (500 and 600 for films below 50\,nm thickness)  controls the film thickness, which was measured by AFM after UV-photolithographic structuring and wet-chemical etching.

The phase purity and orientation of the deposited films were analysed via X-ray diffraction (XRD) with Cu K$\alpha$ radiation; $\theta$-$2\theta$-scans were performed to verify the $c$-axis orientation of deposited films. The phase purity of the deposited Bi-2212 thin films, which can often suffer from the low-$T_\mathrm{c}$ phase ingrowth of Bi$_2$Sr$_2$CuO$_y$, was confirmed.

For the TEM studies, thin epitaxial Bi-2212 film were grown on LAO and LSAT substrates and covered by a Pt-C protective layer before a focused gallium ion beam (Ga-FIB) was used to cut a thin lamella. Short imaging times of only a few minutes were used to not deteriorate the atomic structure, as longer times led to a complete amorphization of the film material.

For the electrical transport measurements, circular contact pads were established by evaporating Ag/Au pads (for optimal adhesion on the Bi-2212 film) through a copper shadow mask. Subsequently, the thin films with thicknesses 46\,nm (42\,nm) were patterned into rectangular-shaped bridges of $63\,\mu$m ($102\,\mu$m) width and $94\,\mu$m ($497\,\mu$m) distance between the voltage probes for sample A (sample B) by UV photolithography and wet chemical etching. Sample B was post-annealed in oxygen after patterning (400$\,^\circ$C for 30\,min in 0.8\,mbar oxygen atmosphere). Electrical contacts in the cryostat were established using silver paste and $50\,\mu$m thick gold wires.

\begin{table}[t!]
\vskip 5mm
  \centering
  \setlength{\tabcolsep}{10pt}
  \begin{tabular}{|l|c|c|c|c|c|c|c|c|c|}
    \hline
     \rowcolor{lightgray} & $T_\mathrm{c,inf}$ & $B_\mathrm{c2}^{\perp}(0)$ & $B_\mathrm{c2}^{\parallel}(0)$ & $\xi_{ab}$ & $\xi_{c}$ & $J_\mathrm{c}^\mathrm{sf}(45\,\mathrm{K})$ & $\lambda_{ab}$ & $\gamma$ & $\kappa$ \\
     \rowcolor{lightgray} & (K) & (T)& (T) & (nm) & (nm) & (MA/cm$^2$) & (nm) & & \\
    \hline
    Sample A & 90 & 106 & 19,000  & 1.8 & 0.01 & 2.89 & $\sim 200$ & 180 & $\sim 113$ \\
    \hline
    Sample B & 88 & 86  & --       & 2.0 & --   & 2.20 & $\lesssim 240$ & --  & $\lesssim 120$ \\
    \hline
  \end{tabular}
  \caption{Summary of the superconducting parameters of very thin Bi-2212 films obtained in this study; Critical temperature $T_\mathrm{c,inf}$ at the inflection point of the transition, upper critical fields with $B$ oriented perpendicular and parallel to the CuO$_2$ layers, $B_\mathrm{c2}^{\perp}(0)$ and $B_\mathrm{c2}^{\parallel}(0)$, in-plane and out-of-plane Ginzburg-Landau coherence lengths at zero temperature $\xi_{ab}$ and $\xi_{c}$, self-field critical current at 45\,K $J_\mathrm{c}^\mathrm{sf}(45\,\mathrm{K})$, in-plane London penetration depth at zero temperature $\lambda_{ab}$, anisotropy parameter $\gamma$, and Ginzburg-Landau parameter $\kappa$. The uncertainties in the evaluation of these parameters are discussed in the text.}
  \label{tab:params}
\end{table}

\subsection*{Transport measurements}

The sample's magnetoresistance $R(T,B)$ as a function of temperature $T$ and magnetic field $B$ oriented perpendicular and parallel to the surface was measured in standard four-probe geometry in a physical properties measurement system (PPMS) using an external constant-current source (Keithley 6221) and a nanovoltmeter (Keithley 2182). The orientation of the magnetic field parallel to the $ab$ direction of the Bi-2212 unit cell was adjusted by taking advantage of the sharp resistance minimum at the vortex lock-in \cite{TACH89} in an applied field at 9\,T. For all orientations of the magnetic field, the transport current was perpendicular to the field. Measurements in zero or low magnetic fields were performed in a cryocooler placed between the pole pieces of an electromagnet.

The critical current was determined under isothermal and constant magnetic field conditions by increasing the current in small steps up to an electric field criterion of $10\,\mu$V/cm (corresponding to 100\,nV for sample A and 500\,nV for sample B) and interpolating between data points. Data at intense currents and low temperatures were excluded when the PPMS indicated a temperature rise due to sample heating by finite contact resistances. The Hall voltage $V_y$ was collected during temperature sweeps at fixed magnetic fields in both polarities $B^+$ and $B^-$ and calculated as $V_y = (V_{B^+} - V_{B^-})/2$. Occasional magnetic field sweeps confirmed the accuracy of this method.

\section*{Data availability}
The data generated and analysed during this study are available from the corresponding author upon reasonable request.

\bibliography{BSCCOv1}

\section*{Acknowledgements}

We are grateful to E.~F. Talantsev for illuminating discussions and to W.~P. Crump for providing their software code. This research was funded in whole, or in part, by the Austrian Science Fund (FWF) Grant-DOI: 10.55776/I4865-N. For the purpose of open access, the authors have applied a CC BY public copyright license to any Author Accepted Manuscript version arising from this submission. B.A. acknowledges  financial support from the Austrian Science Fund, grant I6079. This article is based upon work from COST Actions SuperQuMap CA21144, Hi-SCALE CA19108, and Polytopo CA23134 (European Cooperation in Science and Technology).

\section*{Author contributions}

W.L. and J.D.P. conceived the experiments, S.K. fabricated and characterized the samples, B.A. performed the transport measurements, B.A. and W.L. analysed the results. W.L. drafted the manuscript and all authors contributed to the manuscript and reviewed it.

\section*{Declaration}
\vskip 8 pt
\section*{Competing interests}
The authors declare no competing interests

\section*{Additional information}
{\bf Correspondence} and requests for materials should be addresses to W.L..
\vskip 8 pt
\noindent {\bf Reprints and permissions information} is available at \href{https://www.nature.com/reprints}{www.nature.com/reprints}.
\vskip 8 pt
\noindent {\bf Publisher’s note} Springer Nature remains neutral with regard to jurisdictional claims in published maps and institutional affiliations.
\newpage
\noindent {\bf Open Access} This article is licensed under a Creative Commons Attribution 4.0 International License, which permits use, sharing, adaptation, distribution and reproduction in any medium or format, as long as you give appropriate credit to the original author(s) and the source, provide a link to the Creative Commons licence, and indicate if changes were made. The images or other third party material in this article are included in the article’s Creative Commons licence, unless indicated otherwise in a credit line to the material. If material is not included in the article’s Creative Commons licence and your intended use is not permitted by statutory regulation or exceeds the permitted use, you will need to obtain permission directly from the copyright holder. To view a copy of this licence, visit \href{http://creativecommons.org/licenses/by/4.0/}{http://creativecommons.org/licenses/by/4.0/}.
\vskip 5 pt
\noindent \textcopyright\  The Author(s) 2025

\end{document}